\documentclass[english]{iopart}
\usepackage[T1]{fontenc}
\usepackage[latin9]{inputenc}
\usepackage{color}
\usepackage{amstext}
\usepackage{amssymb}
\usepackage{graphicx}
\usepackage{esint}

\makeatletter

\providecommand{\tabularnewline}{\\}

\usepackage{iopams}
\usepackage{setstack}

\makeatother

\newcommand{\relieve}[1]{#1}

\newcommand{\revise}[1]{#1}
\definecolor{blue}{rgb}{0,0,1}

\usepackage{babel}
\begin{document}

\title[Speeding up pedestrian and granular bottleneck
flows]{A counterintuitive way to speed up pedestrian and granular bottleneck
flows prone to clogging: \emph{Can `more' escape faster? }}

\author{Alexandre NICOLAS$^{1}$, Santiago IB\'A\~NEZ$^{2,3}$, Marcelo
N. KUPERMAN$^{2,3}$, Sebasti\'an BOUZAT$^{2,3}$}

\address{$^{1}$ LPTMS, CNRS, Univ. Paris-Sud, Universit\'e Paris-Saclay,
91405 Orsay, France}

\address{$^{2}$ Consejo Nacional de Investigaciones Cient\'{\i}ficas y T\'ecnicas\\
$^{3}$ Centro At\'omico Bariloche (CNEA) and Instituto Balseiro,
R8400AGP Bariloche, Argentina.}

\begin{abstract}
Dense granular flows through constrictions, as well as competitive pedestrian evacuations,
are hindered by a propensity to form clogs. \revise{We use simulations of model pedestrians and experiments with granular disks to} explore an original strategy to speed up these flows, which consists in
including contact-averse entities in the assembly.
On the basis of a minimal cellular automaton and a continuous agent-based model
for pedestrian evacuation dynamics, we find that the inclusion of polite pedestrians amid a given competitive crowd fails to
reduce the evacuation time when the constriction (the doorway) is acceptably large. This is not surprising, because adding agents
makes the crowd larger. In contrast, when the door is so narrow that it can accommodate at most one or two agents at a time,
our strategy succeeds in substantially curbing long-lived clogs and speeding up the evacuation. A similar effect is seen
experimentally in a vibrated two-dimensional hopper flow with an opening narrower than 3 disk diameters. Indeed, by adding
to the initial collection of neutral disks a large fraction of magnetic ones, interacting repulsively, we observe a shortening of the
time intervals between successive egresses of neutral disks, as reflected by the study of their probability distribution. 
On a more qualitative note, our study suggests that the much discussed analogy between pedestrian flows and granular flows could be extended to some 
behavioural traits of individual pedestrians.
\end{abstract}
\maketitle

\section{Introduction}

Clogs rank high on the list of nightmares of grain silo managers and
crowd managers alike. Every year, grain entrapment still claims the
lives of dozens of workers who crawled inside bins or silos to knock
down clogs and speed up the outpour. For instance, in the United States,
it was at the origin of 11 fatalities and 18 non-fatal injuries in
2016 \cite{issa2017summary}, while in France \cite{BARPI2017accidentologie}
and in Argentina \cite{superintendencia2016informe} fatal accidents
occur every few years. Concurrently, crowd disasters due to massive
rushes towards an exit in situations of emergency hit the headlines
on a yearly basis. 

From a fundamental perspective, recent works have shed light on similarities
in the clogging and unclogging dynamics across a wide variety of systems,
from vibrated granular flows to pedestrian flows through a bottleneck
\cite{Zuriguel2014clogging}. In the study of granular flows through
a silo or a hopper, different strategies have been investigated to
avoid or destroy clogs. On account of the sensitivity of the flow
to the size of the outlet, enlarging the latter is an intuitive and
efficient strategy, but researchers have also studied the effect of
vibrating the setup vertically \cite{mankoc2009role,lozano2012breaking}
or making the exit oscillate in the horizontal plane \cite{to2017flow}.
Placing an obstacle in front of the exit was also found to enhance
the flow in specific conditions \cite{zuriguel2011silo,endo2017obstacle}. The latter
strategy may also be efficient to facilitate the evacuation of living
beings, such as sheep and mice, but the results seem to strongly depend
on the position of the obstacle \cite{zuriguel2016effect} and the
geometry of the setup \cite{lin2017experimental}, and it has not
been demontrated yet that this method also applies to real pedestrian crowds \cite{yano2018effect}.
In contrast, an aspect that is known to considerably affect pedestrian
flow rates is the motivation or eagerness to escape. In a first regime,
a crowd that is more eager to escape will effectively evacuate faster
\cite{nicolas2017pedestrian}. However, beyond a certain level of
competitiveness, further increasing the eagerness to egress results in the
build-up of higher pressure at the bottleneck, due to physical contacts,
which stabilises clogs and thus delays the evacuation \cite{parisi2007faster,Pastor2015experimental}.
This is the well-known `faster-is-slower' effect, first evidenced
in numerical simulations \cite{helbing2000simulating} (note that
similar effects are observed when increasing the effective gravity
in granular hopper flows \cite{Zuriguel2014clogging}). Pedestrian
flow is thus enhanced by maintaining the crowd's competitiveness in-between
these two regimes, where the exit capacity reaches its peak value.
\revise{
Previous numerical studies of evacuation dynamics have also considered crowds with
heterogeneous behaviours \cite{Heliovaara2013patient, parisi2015human,nicolas2016statistical}.
These behaviours are split into two categories: cooperative \emph{vs} non-cooperative\footnote{The behaviours are called `patient' and `impatient'
in \cite{Heliovaara2013patient}, while ant-like \emph{vs} human-like strategies are considered in
\cite{parisi2015human}.}.  Non-cooperative behaviour was found to limit the evacuation efficiency owing to
the formation of clogs and jams, but purely cooperative behaviour was also suboptimal, because of the placidity of the
agents. The fastest evacuations were achieved by mixed crowds, with an optimal mixture of agents.}

In this contribution, we examine a \revise{somehow counter-intuitive} way to accelerate
bottleneck flows of highly competitive crowds and grains prone to
clogging: \revise{Instead of varying the fraction of cooperators within a crowd,
we consider a given homogeneous non-cooperative population
and bring in \emph{more} entities. Na\"ively, one might expect that the evacuation
will then always take longer, because the assembly is larger. However,
given that} clogs are due to force-bearing contacts
at the bottleneck, the injection of entities (pedestrians or grains)
that shun contacts may actually limit these clogs. We show that such dilution among
contact-averse entities can effectively lead to a quicker evacuation
of the original system, even though overall more entities need to
escape. We start by exposing some theoretical arguments to support
the possible occurrence of this paradoxical effect in Sec.~\ref{sec:Theory}.
Then, we test it in a cellular automaton model for pedestrian evacuation
through a bottleneck in Sec.~\ref{sec:Automaton} and a related agent-based
model (operating in continuous space) in Sec.~\ref{sec:Agent}. Finally,
in Sec.~\ref{sec:Granular}, we report on granular hopper flow experiments 
with a mix of magnetic and neutral (non magnetic) disks, in which the magnetic disks 
\relieve{play the role} 
of contact-averse entities.

%  Original
%Finally,
%in Sec.~\ref{sec:Granular}, since an empirical validation with pedestrians
%is too risky, we exploit the tentative analogy between grains and
%pedestrians in some conditions and report on granular hopper flow
%experiments in which cooperative pedestrians are mimicked by repulsive
%magnetic disks. Beyond the analogy with pedestrians, it is worth remarking
%that speeding up hopper flows via the introduction of repulsive particles
%may have interesting applications \emph{per se}.

\section{Theoretical principle\label{sec:Theory}}

The theoretical idea underlying the method proposed is easy to grasp.
Consider an assembly of $N$ discrete bodies (\emph{e.g.}, pedestrians),
$n$ of which are prone to contact (that is, highly competitive in
the case of pedestrians) while the remaining $m=N-n$ are not. Let
$T(n,m)$ be the total time needed for evacuation through a bottleneck
of width $w$. For a large crowd ($N\gg1$), in the absence of macroscopic
segregation, it is fair to approximate $T$ as 
\[
T(n,m)\approx\frac{n+m}{J\left(\frac{m}{m+n}\right)},
\]
where $J(c_{m})$ is the steady-state flow rate for a well-mixed assembly
made of a fraction $c_{m}=\frac{m}{m+n}\in[0,1]$ of contact-averse
bodies.

For crowds, the experimentally observed `faster-is-slower' effect
\cite{Pastor2015experimental} suggests that $J$ may increase with rising $c_m$ near
$c_{m}=0$ \footnote{Videos of the clogs obtained in this limit can be found on the internet, \emph{e.g.} at https://www.youtube.com/watch?v=OCKO0c6lihU}.
Therefore, the minimum evacuation time $T$ at fixed $n$ will not
necessarily be reached for $m=0$. Taking the continuous limit $m\in\mathbb{R}$,
one can write 
\[
\frac{\partial T}{\partial m}\Big|_{n}=\frac{J\left(c_{m}\right)-\left(1-c_{m}\right)\,J^{\prime}\left(c_{m}\right)}{J^{2}\left(c_{m}\right)}.
\]
Any extremum of $T$ reached for  $c_{m}\in(0,1)$ should obey
\begin{eqnarray*}
J\left(c_{m}\right)-\left(1-c_{m}\right)J^{\prime}\left(c_{m}\right) & = & 0.\\
\frac{d}{dc_{m}}\left[\left(1-c_{m}\right)\,J(c_{m})\right] & = & 0
\end{eqnarray*}
It follows that the evacuation time of the contact-prone entities will be extremised
at $c_{m}>0$, i.e., in the presence of other people, if the partial
flow rate $\left(1-c_{m}\right)\,J(c_{m})$ is non-monotonic. 
\revise{For instance, should $J(c_{m})$ be strongly reduced
due to clogs when $c_{m}\to0$, this
condition will be met.} 

This raises two questions. First, beyond its theoretical possibility,
can such reduction of delays at bottlenecks through `dilution' actually
occur in \emph{realistic} settings? Secondly, can the proposed mechanism
be implemented \relieve{concretely for practical applications}? The following sections address the first
question, while the second issue will be touched upon in the conclusion.

\revise{
To address the problem quantitatively, in addition to the \emph{global} evacuation time $T$ and flow rate $J$,
we will inspect the series of exit times $t_i$, with $i \in [1,N]$, and 
compute the time gaps $\tau_{i}=t_{i}-t_{i-1}$ between successive egresses,
where $i\in[2,N]$, as well as the time gaps $\theta_{j}=t_{\sigma(j+1)}-t_{\sigma(j)}$ between egresses of contact-prone entities, 
where $\sigma(j)$ is the egress rank of the $j$-th such entity and $j\in[2,n]$. Note
that the global evacuation time is given by 
\[
T(n,m)=t_{1}+\sum_{i=2}^{N}\tau_{i}=t_{\sigma(1)}+\sum_{j=2}^{n}\theta_{j},
\]
so, for large $n$, the mean evacuation time per \emph{contact-prone} (competitive) entity is given by
$\langle\theta\rangle \simeq \frac{T(n,m)}{n}$. 
}

\section{Cellular automaton model for pedestrian evacuation through a bottleneck\label{sec:Automaton}}

We start by exploring the validity of the idea exposed in the previous
section with a cellular automaton developed by some of us \cite{nicolas2016statistical}.
The model was shown to semi-quantitatively
reproduce the experimental pedestrian-scale dynamics
\cite{Pastor2015experimental} of pedestrian flows through a narrow
door, notably the statistics of time gaps between escapes. Here, we
consider a crowd of $n$ highly competitive (impatient) pedestrians
and $m$ patient agents, all of whom wish to evacuate through a door
of width $L_{d}$. Patient and impatient agents are characterised
by a propensity to cooperate $\Pi$ that is uniformly distributed
in the intervals $(0.8,\,1)$ and $(0,\,0.2)$, respectively.

\subsection{Brief description of the cellular automaton model}

To make the paper self-contained, we briefly recall the main features
of the model. Space is discretised into a square lattice with at most
one agent per site. At each time step, \\
(1) each agent behaves either cooperatively (with probability $\Pi$)
or competitively (with probability $1-\Pi$);\\
(2) all agents select a target site among the four adjacent sites
(von Neumann neighbourhood) plus the current one, with probabilities
that depend on the proximity of the sites to the exit and a strong
preference for empty sites. The probability to select the current
site, i.e., to choose to stay on site, is strongly reduced if the
agent behaves competitively.\\
(3a) if the target site is occupied, the agent just waits;\\
(3b) otherwise, he/she moves to it, unless other agents are competing
for it (in which case no one moves). \\
(4) Following this first round of motion, some sites have been freshly
vacated, which may allow waiting pedestrians to move to their target
site. Steps (3) are iterated until all possibilities of motion have
been exhausted. 

The iterative rule (3) allows the formation of files of moving pedestrians,
without voids. Note that agents cannot move more than once during
a time step and that rule (3b) corresponds to the limit of strong
friction, where any competition for a site is counterproductive.

\subsection{Results}

We consider a crowd of $N=n+m$ agents, where $n=6,000$ will be kept
fixed while $m$ will be varied, and simulate its evacuation through
a door of the width of two agents, i.e., $L_{d}=2$.

\begin{figure}
\begin{centering}
\includegraphics[width=0.48\textwidth]{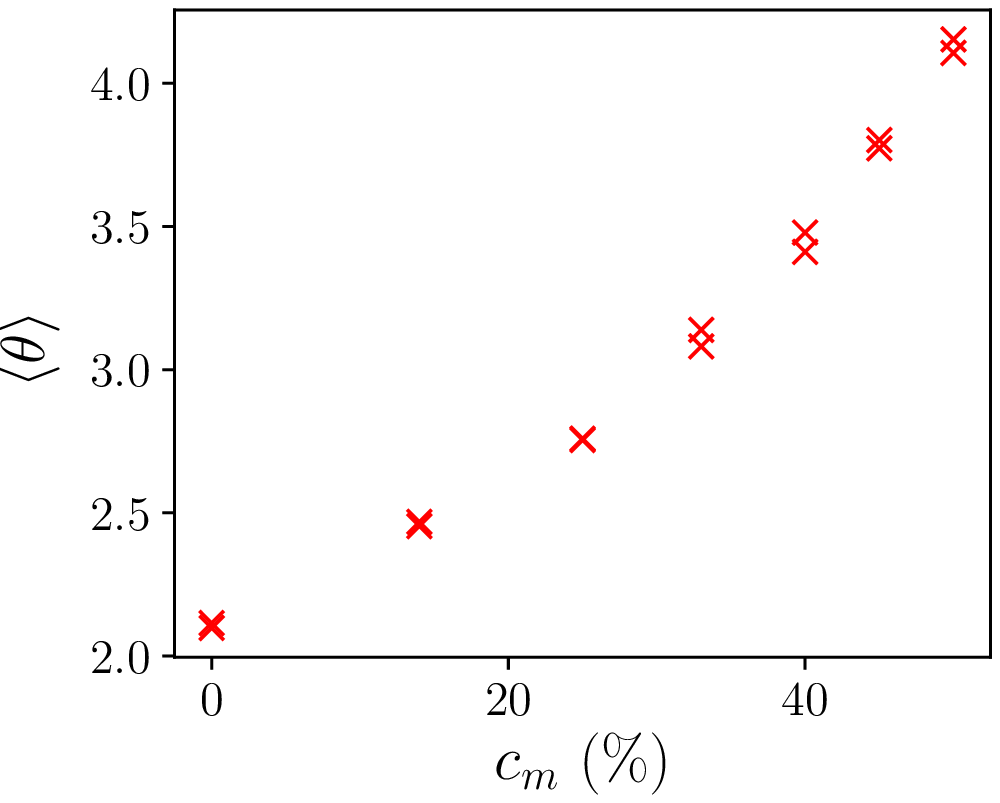}~\includegraphics[width=0.48\textwidth]{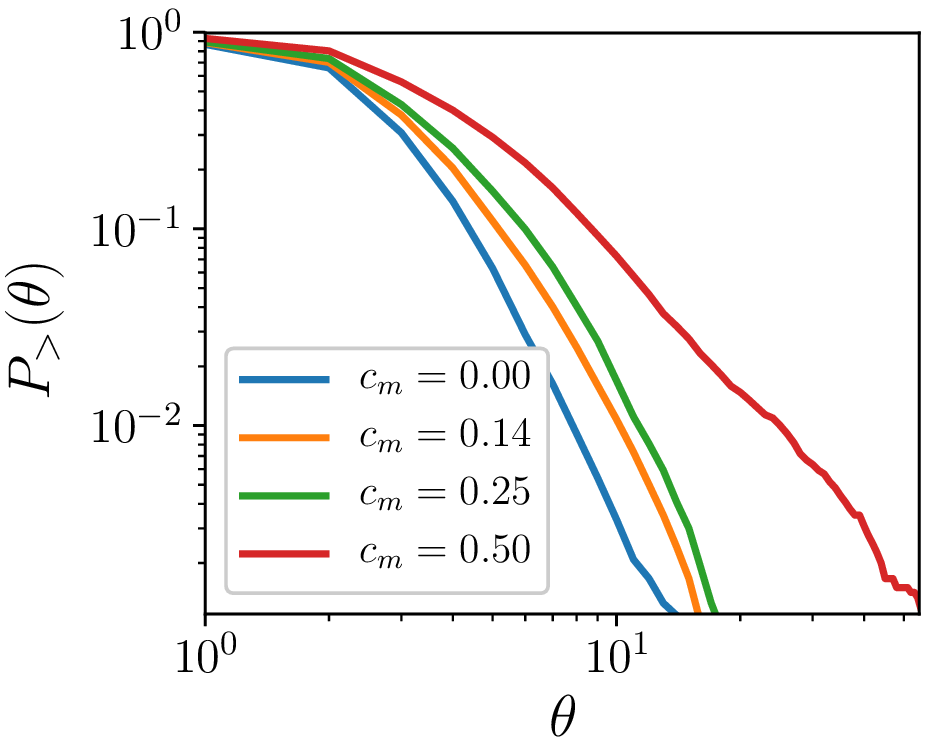}
\par\end{centering}

\caption{\label{fig1}(Colour online) Pedestrian bottleneck
flow through a door of size $L_{d}=2$ \revise{simulated with a cellular automaton.} \textbf{(a) }Variation of
the mean escape time $\langle\theta\rangle=T/n$ per competitive agent
with the fraction of $c_{m}$. Each cross corresponds to one realisation
of an evacuation of $n=6,000$ competitive agents. \textbf{(b)} Survival
function (CCDF) $P_{>}(\theta)$ of time intervals $\theta$ between
\emph{competitive} agents for different fractions $c_{m}$.}
\end{figure}

Figure~\ref{fig1}(a) shows that the mean evacuation time per competitive agent $\langle\theta\rangle$ increases
with $m$ \revise{for $L_d=2$.} This implies that the simulated evacuation lasts longer
as more patient agents are inserted. Thus, the dilution strategy fails
to substantially fluidise the flow. It may be worth mentioning that
we had come to a similar conclusion in previously published controlled
experiments in which a mixed crowd made of polite and selfish agents
evacuated through a narrow door. Indeed, the outflow was found to
slow down as the fraction $c_{m}$ of polite participants in the crowd
increased \cite{nicolas2017pedestrian}. \revise{This observation should
be put in parallel with the absence of long-lasting clogs in the evacuation of the
purely selfish crowd ($c_m=0$): For the chosen door width, as a result of the safety prescriptions, the participants
did not behave competitively enough to make the evacuation dynamics strongly
intermittent.}

In our simulations, the delay induced by the addition of $m$ patient
agents in the crowd is confirmed by plotting the complementary cumulative
distribution function (CCDF) $P_{>}(\theta)=\int_{\theta}^{\infty}p(\theta^{\prime})\,d\theta^{\prime}$,
where $p(\theta)$ is the probability density function of time gaps
between \emph{competitive} agents. Note that $\langle\theta\rangle$
can be deduced from the CCDF via $\langle\theta\rangle=\int_{0}^{\infty}P_{>}(\theta^{\prime})\,d\theta^{\prime}$.
As can be seen in Fig.~\ref{fig1}(b), larger time gaps $\theta$
become more frequent as the fraction of patient agents $c_{m}=\frac{m}{N}$
increases.

\begin{figure}
\begin{centering}
\includegraphics[width=0.48\textwidth]{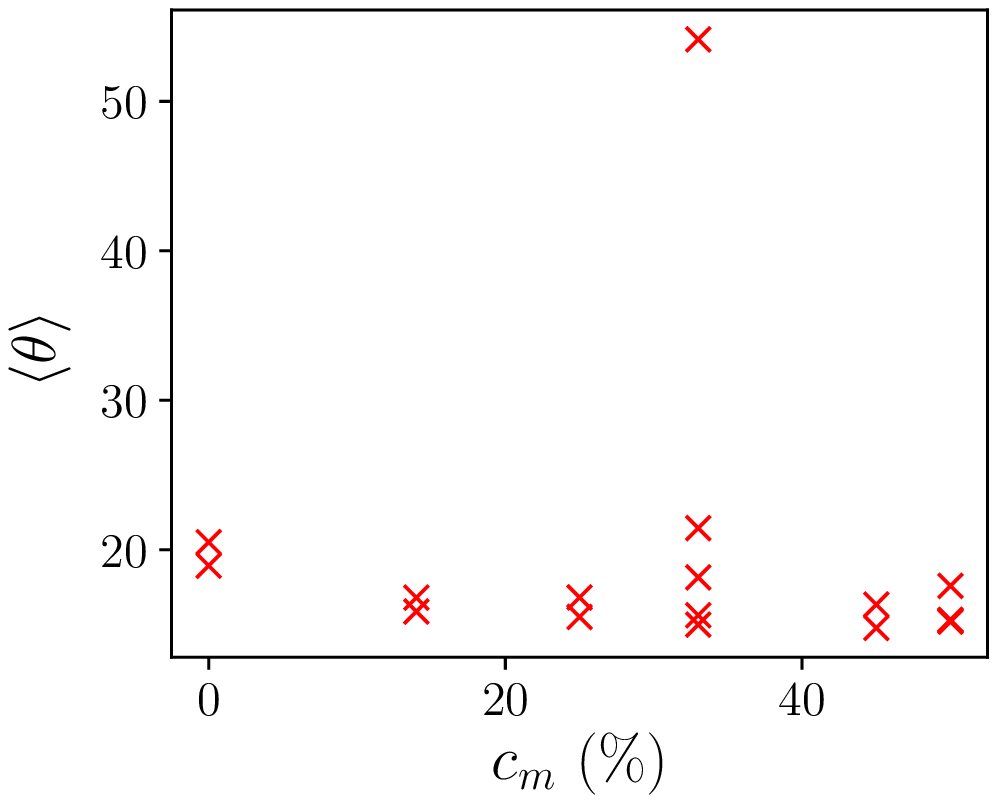}~\includegraphics[width=0.48\textwidth]{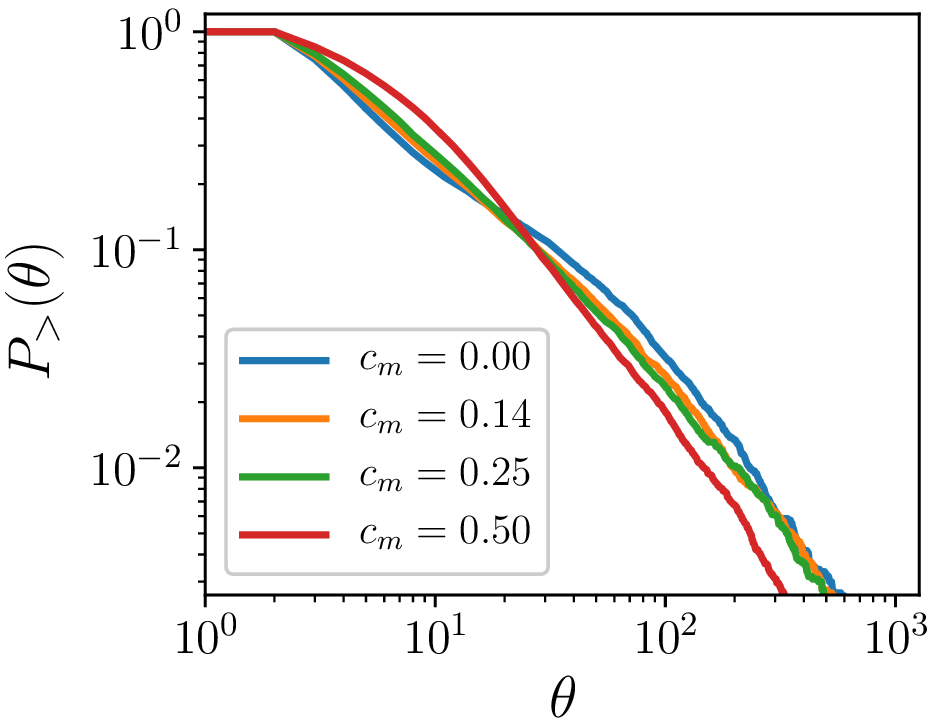}
\par\end{centering}

\caption{\label{fig2}(Colour online) Pedestrian bottleneck
flow through a door of size $L_{d}=1$ \revise{simulated with a cellular automaton.} \textbf{(a) }Variation of
$\langle\theta\rangle$ with $c_{m}$. Each cross corresponds to one
realisation of an evacuation of $n=6,000$ competitive agents. \textbf{(b)}
Survival function $P_{>}(\theta)$ for different fractions $c_{m}$.}
\end{figure}

In light of the failure of the fluidisation strategy
for $L_{d}=2$, we consider an even narrower door, $L_{d}=1$, which
should lead to even more clogging. This narrowing considerably affects
the results. The evacuation time $\langle\theta\rangle$ per competitive
agent then seems to be reduced as patient agents are introduced in
the crowd and reaches its minimal value for a fraction $c_{m}>0$
{[}Fig.~\ref{fig2}(a){]}. However, despite the large crowd ($N>6,000)$,
very large fluctuations occur between realisations under the same
conditions, typical of clogged flows, which hampers a clear appraisal
of the evolution of $\langle\theta\rangle$ with $c_{m}$. Here, the seemingly erratic outliers at surprisingly
large values of $\langle\theta\rangle$ originate in the 
fortuitous encounter of extremely competitive agents ($\Pi \to 0$) at the door. Therefore,
we turn to an inspection of the CCDF $P_{>}(\theta)$ in Fig.~\ref{fig2}(b).
While the addition of patient agents (larger $c_{m}$) still increases
the overall frequency of moderate to long time gaps $\theta$ between
competitors ($\theta>10$), it results in rarer very long clogs (leading
to $\theta>30$), as the doorway is less likely to be blocked by an
encounter between extremely competitive agents. Even though such clogs
are rare events, they are so long that they significantly reduce the
flow rate. The strategy of introduction of patient agents thus succeeds
in fluidising the competitors' flow via the alleviation of these long
clogs.

\section{Agent-based model for pedestrian dynamics\label{sec:Agent}}

\subsection{Brief description of the model}

In the previous section, flow enhancement through
the incorporation of polite agents was proven possible within a lattice-based model, which had been 
calibrated for the evacuation of \emph{homogeneous}
crowds. To test the robustness of these predictions, we now investigate
a second model for pedestrian evacuation dynamics. This model, introduced
in \cite{dossetti2017behavioral}, operates
in discrete time and continuous space, and it also considers the influence
of different individual attitudes. Agents are modelled as hard disks
of diameter $d=1$ and continuously make decisions about how to move
in a changing environment. When possible, they step into free space
but, 
\relieve{if little space is available, they adopt a
behaviour that is either patient (cooperative strategy) or
impatient (competitive strategy).
Patient agents will only attempt a move if some distance $\mu=\mu_{m}$,
with $0<\mu_m<1$ is available in the desired direction, whereas their impatient
counterparts will try to move as soon as they see a distance $\mu=\mu_{n}$ in this direction, with $0 <\mu_n \ll \mu_m$.
We will vary $\mu_n$ while keeping $\mu_m=0.6$ fixed. The desired direction is drawn randomly within a
window of angular width $\eta$ around the direction towards the exit or, if it is obstructed, one of the two perpendicular directions.}

\relieve{
More precisely, at the beginning of the evacuation, $N$ pedestrians are randomly
distributed (without overlaps) in a square room of size $L$, with
an exit in the middle of one of the walls. At each time step,\\
(1) every agent \emph{considers} a move towards the
exit, with a small random angular fluctuation of 
standard deviation $\eta=\pi/8$ around the direction of the exit;\\
(2a) if the gap distance $D_\parallel$ to the nearest obstacle (e.g., another agent) along the desired direction is larger than $\mu$ (which depends on the strategy), the agent \emph{attempts} a forward step of size $\min(1,D_\parallel)$;\\
(2b) otherwise, the agent \emph{considers} a lateral move to the right or to the left, within $\eta$ of the perpendicular to the exit direction; \\
(2c) if the gap distance $D_\perp$ to the nearest obstacle along that direction is larger than $\mu$, the agent \emph{attempts} a lateral step of size $\min(1,D_\perp)$;\\
(3) Once all individuals have defined their desired moves, conflicts
may arise if these moves lead to collisions; they
are settled by randomly selecting a winner among the rivals and letting her move. \\
(4) all agents move to their new positions.\\
}
The update dynamics are synchronous.

\subsection{Results}

\begin{figure}
\noindent \begin{centering}
\includegraphics[width=0.8\textwidth]{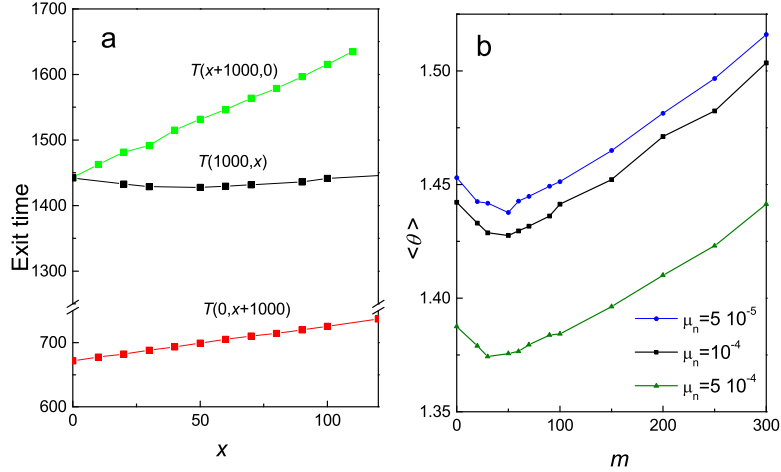}
\par\end{centering}

\caption{\label{fig:evac_times_cont_model} 
\relieve{ Evacuations of $N$ 
pedestrians through a door of size $L_d=2d$ in the agent-based model. 
(a) Global evacuation times for a purely competitive crowd ($N=n=1000+x$), a purely cooperative one 
($N=m=1000+x$) and a mixed one, ($N=n+m$, $n=1000$ and $m=x$). 
Here $\mu_n=10^{-4}$ and $\mu_m=0.6$
(b) Evacuation times per competitive agent for a mixed crowd composed of 1,000 competitive agents and $m$ cooperative ones, as
a function of $m$, for distinct minimal distances $\mu_n$. }}
\end{figure}

\revise{
Let us first consider a purely competitive population with $N=n=1000$ agents (while $m=0$), and on the other hand 
a purely cooperative population with $N=m=1000$ ($n=0$).
Figure~\ref{fig:evac_times_cont_model}(a) shows that, as expected, 
the addition of $x$ agents of the same type as the original ones to either of these
populations lengthens the evacuation: $T(1000+x,0)$ and $T(0,1000+x)$ grow 
monotonically with $x$. In contrast, when $x$ {\em cooperative} agents are inserted 
amidst the {\em competitive} population, the evacuation time $T(1000,x)$ is found to decrease for small $x$
and reaches a minimum close to $x=m^\star\approx50$. For $x>m^\star$, the delay associated with the egress 
of the cooperators is no longer compensated by the higher flow rate, and the evacuation time increases with $x$.
The effect is better shown in Fig.~\ref{fig:evac_times_cont_model}(b), where we plot the evacuation time per competitive 
agent $\langle \theta \rangle \equiv T(n,m)/n$ as a function of $m$ for fixed $n=1000$. 
The effect holds for the three different values of the minimal distance $\mu_n$ that we have tested.
}

\begin{figure}
\noindent \begin{centering}
\includegraphics[width=0.8\textwidth]{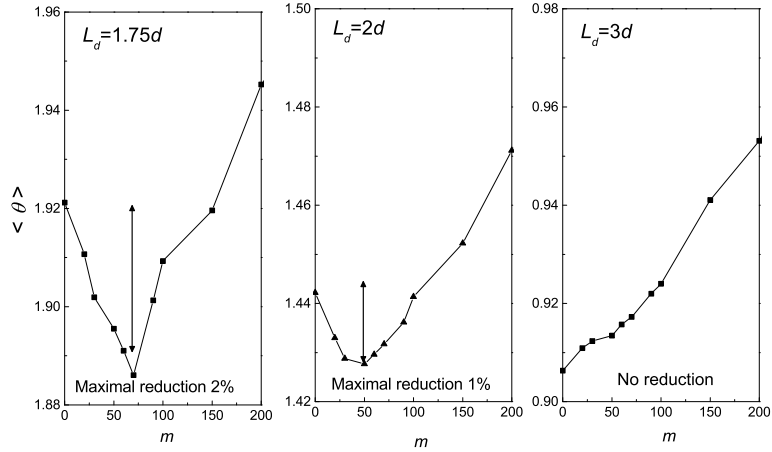}
\par\end{centering}

\caption{\label{fig:evac_times_cont_model2}
Dependence on the door width $L_d$ of the evacuation times per competitive agent $\langle \theta \rangle \simeq T/n$ for a crowd
composed of 1,000 competitive agents \relieve{($\mu_n=10^{-4}$)} and $m$ cooperative ones, as
a function of $m$, in the agent-based model. }
\end{figure}

Figure~\ref{fig:evac_times_cont_model2} shows how the fluidising
effect of cooperators depends on the door width $L_{d}$. The effect
disappears for a door wider than 3 agents, which is rather similar
to our findings with the cellular automaton (where the effect disappeared
for a door width $L_{d}=2$).
Conversely, the peak reduction in evacuation time, relative to the situation
without cooperators, is enhanced from $1\%$ to $2\%$ (approximately) when the door
is reduced from $L_{d}=2$ to $1.75$. The main reason for this
trend is the strong increase in the propensity for clogs when the
door narrows down [note, in particular, the related increase for $m=0$].

\begin{figure}
\noindent \begin{centering}
\includegraphics[width=0.7\textwidth]{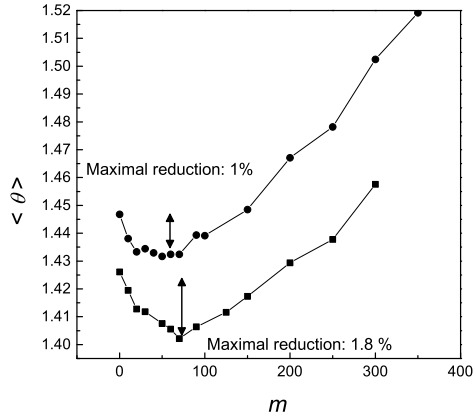}
\par\end{centering}

\caption{\label{fig:evac_times_cont_model3} Influence of the number of competitive agents $n$ ($\mu_n=10^{-4}$)
on the evacuation time  per competitive agent $\langle \theta \rangle \simeq T/n$, shown as a function of $m$ for $n=1,000$ (circles) and $n=1,200$ (squares). 
The door width is $L_d=2d$.}
\end{figure}

Finally, the dependence of $\langle \theta \rangle \simeq T(n,m)/n$ on the number $n$ of competitive
agents to be evacuated is studied in Fig.~\ref{fig:evac_times_cont_model3}.
The relative reduction in evacuation time seems to be enhanced for
larger crowds (larger $n$).

All in all, the agent-based model corroborates the results obtained with the cellular automaton.

\section{Hopper flow experiments with magnetic and non-magnetic disks\label{sec:Granular}}

Having validated the proposed speeding-up method in two distinct models for 
pedestrian dynamics, we now test it in a granular hopper flow experiment.

We studied the flow of mixed magnetic and neutral (non magnetic) disks through a two 
dimensional (2D) vibrated hopper, sketched in Fig.~\ref{figexperim}. The setup was 
previously used in \cite{nicolas2016influence}. The magnetic disks have parallel magnetic 
moments so that they repel each other and thus act as contact-averse entities;
\revise{their magnetic interactions become very small (compared to friction) for distances larger than 1 disk diameter and practically
negligible beyond 3 or 4 diameters.} 
In contrast, neutral disks interact purely mechanically and have a stronger 
tendency to clogging.

As a matter of fact, the study of the hopper flow of magnetic and
non-magnetic disks is a topic worthy of interest \emph{per se}. Surprisingly,
these systems have never been investigated. Some pioneering studies
bore on magnetic grains in a silo \cite{lumay2008controlled}, or
the hopper flow of only magnetic disks \cite{lumay2015flow,hernandez2017discharge}.
In the latter case, however, no vibrations were applied to the setup
and permanent clogs were observed.

Beyond the intrinsic relevance of the granular system, it is interesting to recall 
the analogy that has been pointed out between granular hopper flows and pedestrian 
evacuations \cite{Zuriguel2014clogging,nicolas2016influence}. In fact, in our experiment, magnetic disks mimick 
patient or polite pedestrians (due to their tendency to avoid frictional contacts), while 
neutral disks resemble competitive pedestrians. A subtle difference between the 
granular and pedestrian systems is that magnetic disks will not be repelled by 
neutral ones. Therefore, in some sense, they are less cooperative than polite pedestrians.
Apart from this subtlety, the analogy between pedestrian and granular flows gains practical relevance if one calls to mind
the risk associated with
conducting pedestrian experiments under conditions that favour
large delays at bottlenecks such as those observed in highly competitive 
crowds \footnote{Examples include some emergency evacuations \cite{nytimes1864chile} as well as the huge clog
formed at the gate of a stadium in the 2013 running of the bulls in Pamplona.}.

\subsection*{Experimental setup and data analysis}

We start by describing the setup in detail. Magnetic disks
are about $1\,\mathrm{mm}$-thick and consist of a circular plastic
cap of $13\,\mathrm{mm}$ of diameter, mounted on a commercial Neodyme
magnet of smaller diameter; they weigh $0.55\mathrm{g}\pm0.01\mathrm{g}$.
For the neutral disks, the magnet is replaced by a bronze washer and
the total weight is $0.58\mathrm{g}\pm0.01\mathrm{g}$\footnote{The slightly smaller weight of the magnetic disks
may be thought of as an additional sign of cooperativeness, as it
results in a slightly smaller sliding speed (on average) than that
of the neutral disks, in the absence of collective interactions.}.

\begin{figure}
\noindent \centering{}\includegraphics[width=0.9\textwidth]{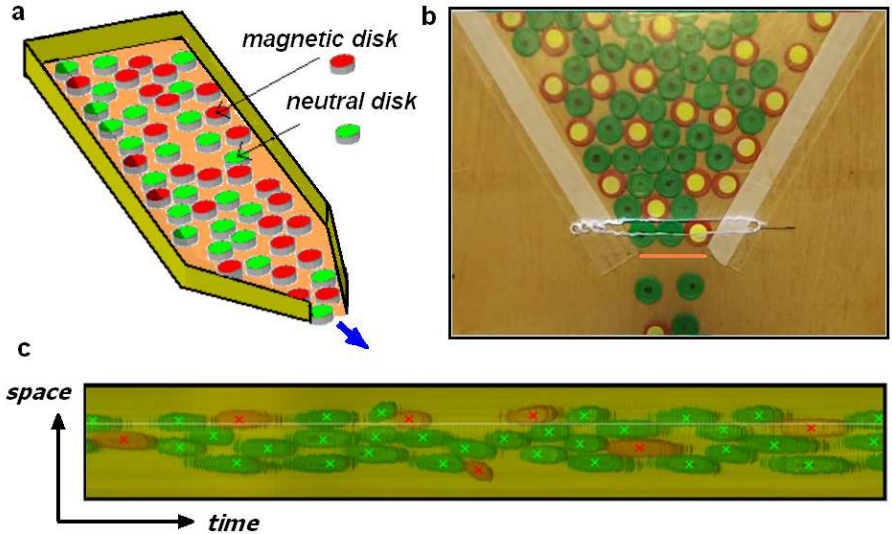}
\caption{(Color online) Experimental setup and video analysis. (a) Sketch of
the setup. (b) Snapshot taken during the discharge of the hopper,
close to the aperture.
 (c) Time frame obtained by stitching lines
of pixels extracted from successive video frames \relieve{(this line is marked in orange in panel b)}, as described in the main text,
and manual tagging of the image.}
\label{figexperim} 
\end{figure}

During the experiments, disks are confined between a tilted chipwood
panel (at the bottom) and a transparent plastic sheet. Thin plastic
bars between the panel and the sheet define a hopper geometry by delimiting
a 2D funnel-like region, with an aperture of width $w=$3 cm \revise{(i.e., 2.3 disk diameters)} and an
opening angle of $58^{\circ}$ (see Fig.~\ref{figexperim}).

The setup is subjected to \revise{(in-plane)} vibrations, \revise{with an amplitude of about 3-6~mm and a frequency of around 10~Hz.} To this end, a motor equipped
with an unbalanced helix was tied underneath the panel. The latter
was carefully polished and varnished to reduce the friction coefficient
and make it as homogeneous as possible. The static friction coefficient
between disks and panel was about $\mu_{s}\simeq0.32$, which corresponds
to an angle of friction $\psi=\tan^{-1}(\mu_{s})\simeq18^{\circ}$;
the plane of the hopper is tilted by an angle roughly equal to (but
slightly smaller) than, $\mu_{s}$. Still, since vibrations tend to suppress
friction, sliding is possible. At the beginning
of each experiment, $n$ neutral (non-magnetic) disks and $m$ magnetic
ones (with typically $N\equiv m+n$ of order 300) are inserted from
the top of the funnel, in random order, and the setup is vibrated
for ten seconds before the start. The error in the number of inserted
disks is generally lower than $2\%$, i.e., $1$-$6$ disks
out of $300$. 

Experiments are filmed with a 60~Hz camera placed above the aperture 
\footnote{Videos are provided as Supplemental Material; in particular, they show how magnetic interactions
can help to break the clogging arches.}.
Egresses of disks through the aperture are detected manually with
the help of a computer-aided routine (similar to that used in \cite{nicolas2016influence,nicolas2017pedestrian},
following \cite{Garcimartin2014experimental,Pastor2015experimental}),
which constructs time frames of escapes by extracting a line of pixels
just past the exit from every frame of the video and stitching these
lines together (an example of such a time frame is shown in Fig. \ref{figexperim}(c)).
The analysis gives access to the series of exit times $t_{i}$, $i\in[1,N]$,
\revise{and to the time gaps $\tau_i$ and time gaps $\theta_i$ between the egresses of neutral disks.}

\begin{figure}
\noindent \centering{}\includegraphics[width=0.9\textwidth]{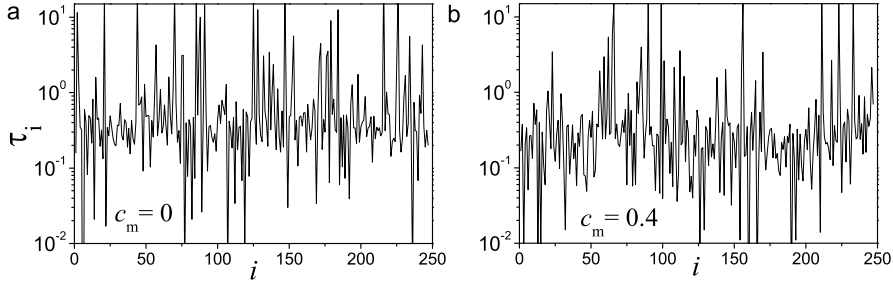}
\caption{Time series of the time gaps $\tau_{i}$ between successive egresses,
labelled by their order of egress $i$, for (a) $c_{m}=0$ and (b)
$c_{m}=0.4$. 
\relieve{Note that the time gaps $\tau_{i}$ which reach the upper bound of the window (15~s) correspond
to long-lived clogs that were eventually destroyed manually.}}
\label{fig-fichas-dt} 
\end{figure}

The granular flows observed experimentally display
highly intermittent dynamics, as shown in~Fig.~\ref{fig-fichas-dt},
with broadly distributed time gaps $\tau$. In particular, long halts
due to clogs are observed. To avoid excessive delays, we artificially
destroy clogs of duration $\tau>15\,\mathrm{s}$, by introducing a
plastic bar in the setup. This procedure does not affect the distribution
of time gaps $p(\tau)$, nor the CCDF $P_{>}(\tau)=\int_{\tau}^{\infty}p(\tau^{\prime})\,d\tau^{\prime}$,
for $\tau<15\,\mathrm{s}$, but it has an impact on the global flow
rate $J = \left(\int_{0}^{\infty}P_{>}\left(\tau\right)\,d\tau\right)^{-1}.$
Accordingly, we define the \emph{capped }time
gaps between egresses of disks, $\overline{\tau_{i}}=\min(15\,\mathrm{s},\,\tau_{i})$,
or neutral disks only, $\overline{\theta_{j}}=\min(15\,\mathrm{s},\,\theta_{j})$. The corresponding \emph{capped}
global and neutral flow rates read
\begin{equation}
\overline{J}=\frac{N-1}{\sum_{i=2}^{n}\overline{\tau_{i}}}\text{ and }\overline{J_{n}}=\frac{n-1}{\sum_{j=2}^{n}\overline{\theta_{j}}}.\label{offflux-1}
\end{equation}
These capped flow rates are unaffected by our manual destructions
of clogs and, were there no clogs that last longer than 15~s, they would match
their \emph{bona fide} counterparts, i.e., $\overline{J}=J$.

\subsection*{Stationarity of the flow}

To examine the stationarity of the flow, we compare the CCDF $P_{>}\left(\tau\right)$
computed for subparts of each experiment, namely, for the first, second,
and third $N/4$ grains, where $N=n+m$ is the total number of grains.
For experiments with $c_{m}<0.5$, no significant difference is seen
between these moving averages. If one discards the last 15\% of grains,
the flow can be regarded as stationary, even though the number of
grains in the hopper varies during the experiment. 
Consistently with Beverloo's law for non-vibrated systems
\cite{beverloo1961flow}, the flow rate is indeed independent of the
column height (as long as it is a couple of times larger than the
width). In contrast, for $c_{m}=1$, the flow is smoother (less intermittent);
it is strongly non-stationary and monotonically decreases with time,
as the height of the granular layer declines. Indeed, (frictional)
force chains are absent in this purely magnetic case, and the pressure
at the bottom thus continuously decreases during the experiment. In
the following, we restrict our attention to situations with $c_{m}<0.5$,
in which quasi-stationarity is achieved. For the same reason, the
last 50-52 disks of each realisation will be discarded.

\subsection*{Analysis of the results}

\begin{figure}
\noindent \centering{}\includegraphics[width=0.9\textwidth]{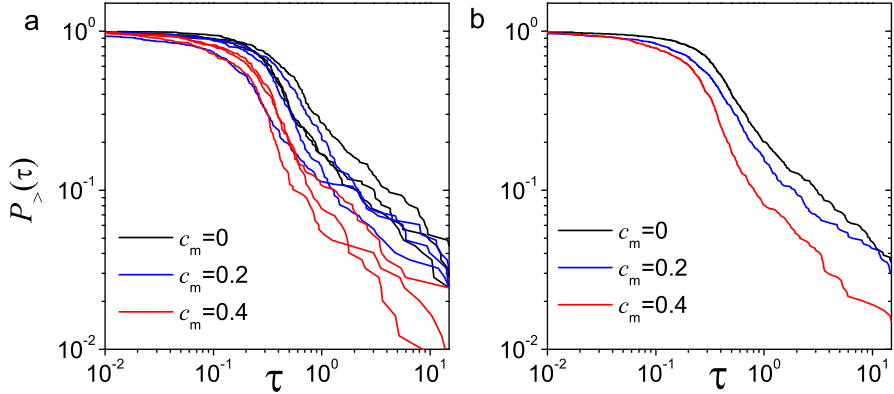}
\caption{(Colour online) Survival functions of the time
gaps $\tau$ between successive egresses \textbf{(a)} for every independent
realisation. \textbf{(b)} for the aggregated data, grouped
according to the fraction $c_{m}$.}
\label{fig-fichas-pcum} 
\end{figure}

Figure~\ref{fig-fichas-pcum}(a) presents CCDFs $P_>(\tau)$ of all time gaps
$\tau$ (irrespective of the nature of the disks) for the 3 realisations
performed for each fraction of magnetic disks: $c_{m}=0$, $c_{m}=0.2$,
and $c_{m}=0.4$. The experimental data are noisy, but what clearly
appears above the noise level is that the CCDFs for $c_{m}=0.4$ decay
faster than those for $c_{m}=0.2$, which decay faster than those
at $c_{m}=0$. The difference appears more clearly if data from realisations
under the same conditions are aggregated [Fig.~\ref{fig-fichas-pcum}(b)]. 

\begin{figure}
\noindent \centering{}\includegraphics[width=0.9\textwidth]{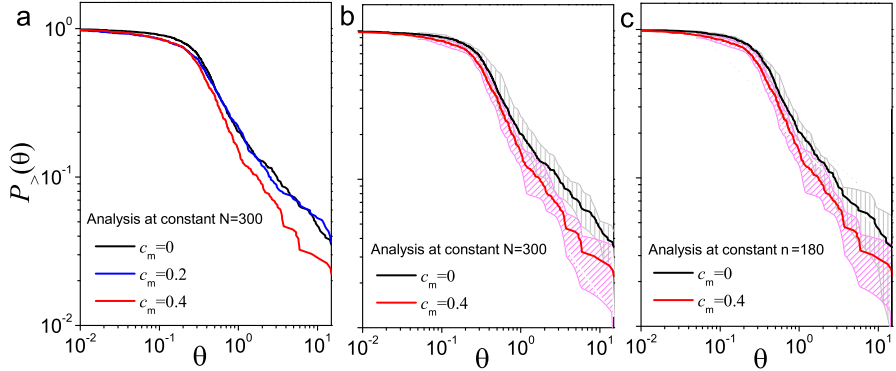}
\caption{(Colour online) Survival functions of time gaps $\theta$ between the egresses of consecutive
\emph{neutral} disks. (Data corresponding to realisations under the same conditions have been aggregated.)
 \textbf{(a)} For the 3 fractions $c_m$ of magnetic disks under study.
\textbf{(b)} For $c_m=0$ and $c_m=0.4$, with envelopes (hatched regions)
\relieve{stretching from the lowest value of $P_>(\theta)$ obtained over the set of realisations to the largest one, for every $\theta$ (three realisations for
each condition).}
\textbf{(c)} Same as panel (b), but this time the analysis is performed over the time window associated with
the discharge of a fixed number of neutral disks, $n=180$, instead of a fixed total number of disks, $N=300$.}
\label{fig-fichas-Penvolv} 
\end{figure}

However, we are not interested in the total flow rate, but only in
enhancing the flow of the neutral disks. Therefore, we now discard
the magnetic disks and consider the time gaps $\theta_{n}$ between
consecutive neutral disks; the CCDFs of the corresponding aggregated
data are plotted in Fig.~\ref{fig-fichas-Penvolv}(a)
\relieve{with envelopes containing all values of $P_>$ found in the different realizations (from the lowest one to the largest one).}
 One can see that the neutral CCDFs decay faster for $c_{m}=0.4$
than for $c_{m}=0$; results for $c_{m}=0.2$ (\emph{not shown}) are
virtually indistinguishable from those obtained for $c_{m}=0$, within
the experimental error bars. Thus, introducing 40\% of magnetic grains
reduces the time gaps between egresses of the neutral grains and,
as a consequence, speeds up the outflow of the latter. This confirms
the validity of our fluidisation strategy in granular hopper flows.
It is interesting to note that the fraction $c_{m}$ of magnetic (`polite')
particles leading to neutral flow enhancement is larger than what
we found in our pedestrian crowd models (see previous sections). This
observation may be associated with our remark that magnetic disks
are `less polite' than their pedestrian counterparts, because they
do not spontaneously avoid contact with neutral disks.

These results were obtained by comparing experiments at different
fractions $c_{m}$ in which the same total number of disks, $N$,
is discharged. Similar conclusions are reached if
one chooses to compare segments of the experiments corresponding to
the discharge of a given number of neutral disks $n$ (while discarding
the initial and final transients, i.e., the first 70 grains
and the last 50 ones, approximately). The neutral CCDFs thus obtained
are plotted in Fig.~\ref{fig-fichas-Penvolv}(b).

The capped neutral flow rates $\overline{J_{n}}$ shown in Table \ref{table}
confirm the significant speed-up of the neutral flow $\overline{J_{n}}(c_{m})$
for $c_{m}=0.4$, \emph{viz., }$\overline{J_{n}}(0.4)>\overline{J_{n}}(0.2)\approx\overline{J_{n}}(0)$,
both at fixed $N$ and at fixed $n$.

\begin{table}
\begin{centering}
\begin{tabular}{cc}
%Experiments at constant $N=300$ \newline

\begin{tabular}{c}
Experiments at constant $N=300$ \tabularnewline
\begin{tabular}{|c|c|c|c|}
\hline 
$c_{m}$  & $0$  & $0.2$  & $0.4$ \tabularnewline
\hline 
$\overline{J}$ & $0.67$  & $0.86$  & $1.46$ \tabularnewline
$\overline{J_{n}}$ & $0.67$  & $0.70$  & $0.98$ \tabularnewline
\hline 
$\langle \overline{\tau} \rangle$ & $1.49$  & $1.16$  & $0.68$ \tabularnewline
$\langle  \overline{\theta} \rangle$ & $1.49$  & $1.42$  & $1.02$ \tabularnewline
\hline 
\end{tabular}
\end{tabular}
%\tabularnewline
%\tabularnewline
& 
\begin{tabular}{c}
Experiments at constant $n=180$ \tabularnewline
\begin{tabular}{|c|c|c|c|}
\hline 
$c_{m}$  & $0$  & $0.2$  & $0.4$ \tabularnewline
\hline 
$\overline{J}$ & $0.69$  & $0.80$  & $1.46$ \tabularnewline
$\overline{J_{n}}$ & $0.69$  & $0.67$  & $0.98$ \tabularnewline
\hline 
$\langle \overline{\tau} \rangle$ & $1.45$  & $1.25$  & $0.68$ \tabularnewline
$\langle \overline{\theta} \rangle$ & $1.45$  & $1.49$  & $1.02$ \tabularnewline
\hline 
\end{tabular}
\end{tabular}
\end{tabular}
%\par
\end{centering}

\caption{\label{table} Capped total and neutral flow rates, $\overline{J}$
and $\overline{J_{n}}$, for different fractions $c_{m}$.  For the sake of exhaustiveness, we also report the
associated mean (capped) time intervals between egresses of disks ($\langle \overline{\tau} \rangle$) or neutral disks ($\langle \overline{\theta} \rangle$).}
\end{table}

\section{Summary and outlook}

In summary, we have studied a strategy to enhance the efficiency of pedestrian evacuations
through a narrow door, prone to clogging, by introducing contact-averse
particles in the system. This strategy is somewhat counterintuitive
insofar as more agents then need to escape to complete the evacuation of
the original system. Still, using two distinct models
for the evacuation dynamics of pedestrian crowds, one based on a lattice
and the other one operating in continuous space, we found that the
positive effect of the reduction of long clogs (caused by the encounter
of competitive agents) could prevail over the negative effect due to
the increase of the crowd size, when the door is very narrow. Moderate
reductions in the evacuation times were then observed. On the other
hand, for wider doors, the strategy is not operational: It heavily
relies on the propensity to clogging displayed by the original system.

On the basis of the established similarities between competitive pedestrian
evacuations and granular hopper flows, we then turned to an experimental
test of the proposed strategy. The role of contact-averse agents was
played by magnetic disks, which interact repulsively with each other,
while the original system was made of neutral disks flowing through
a two-dimensional hopper subjected to vibrations. When a fraction
of around 40\% of magnetic disks was introduced, we observed a faster
decay of the distribution of time gaps between consecutive egresses
of neutral disks, hinting at an effective enhancement of the flow
of the original system. This effect was above the level of experimental
noise and robust to variations in the total number of disks.

Having summarised our findings and shown the technical validity of
the proposed strategy, we now wonder about their practical relevance.
Regarding granular systems, one of the drawbacks of the method is that 
it cannot directly be transposed to three dimensions, where
magnetic interactions would not be only repulsive. But one could contemplate
the incorporation of soft frictionless particles, which are much
less prone to clogging than their hard frictional counterparts \cite{hong2017clogging}, instead of
magnetic disks.
Turning to crowds, \revise{it should be borne in mind that our study is
based on simplistic models: Any extrapolation to real crowds, though sensible,
is speculative. Besides, in practice it is not possible to introduce polite
pedestrians at the last moment, should an emergency occur. Nonetheless, our findings tentatively suggest
that the presence of a small number of attendants trained to keep calm during emergencies (such as safety officers), despite making 
the crowd slightly larger,
might be beneficial in terms of evacuation efficiency \emph{even if} these agents do not modify the behaviour of the rest of the
attendants.
}
\\

\end{document}